\documentclass[notitlepage,twocolumn]{article}
\usepackage{booktabs} 
\usepackage{url}
\usepackage{hyperref}
\usepackage{titling}
\usepackage{blindtext}
\usepackage{graphicx}

\begin{document}
\title{Data Pallets: Containerizing Storage For Reproducibility and Traceability}

\author{
Jay Lofstead\\Sandia National Labs
\thanks{This manuscript has been authored by National Technology and Engineering Solutions of Sandia, LLC. under Contract No. DE-NA0003525 with the U.S. Department of Energy/National Nuclear Security Administration.  The United States Government retains and the publisher, by accepting the article for publication, acknowledges that the United States Government retains a non-exclusive, paid-up, irrevocable, world-wide license to publish or reproduce the published form of this manuscript, or allow others to do so, for United States Government purposes.}
\and
Joshua Baker\\Sandia National Labs
\and
Andrew Younge\\Sandia National Labs
}

\maketitle

    \begin{abstract}
    Trusting simulation output is crucial for Sandia's mission objectives. We rely on these simulations to perform our high-consequence mission tasks given national treaty obligations. Other science and modeling applications, while they may have high-consequence results, still require the strongest levels of trust to enable using the result as the foundation for both practical applications and future research. To this end, the computing community has developed workflow and provenance systems to aid in both automating simulation and modeling execution as well as determining exactly how was some output was created so that conclusions can be drawn from the data.
    
    Current approaches for workflows and provenance systems are all at the user level and have little to no system level support making them fragile, difficult to use, and incomplete solutions. The introduction of container technology is a first step towards encapsulating and tracking artifacts used in creating data and resulting insights, but their current implementation is focused solely on making it easy to deploy an application in an isolated ``sandbox'' and maintaining a strictly read-only mode to avoid any potential changes to the application. All storage activities are still using the system-level shared storage.
    
    This project explores extending the container concept to include storage as a new container type we call \emph{data pallets}. Data Pallets are potentially writeable, auto generated by the system based on IO activities, and usable as a way to link the contained data back to the application and input deck used to create it. 
    \end{abstract}

\section{Introduction}\label{sec:intro}
The existing paradigm for data management and traceability relies on external workflow management tools, such as the Sandia Analysis Workbench (SAW), to track artifacts. Unfortunately, these approaches have been problematic because the workflow engines do not manage the artifacts themselves as much as try to keep track of them. If a user or the system migrates data, then SAW has to figure out what happened. When running Modeling and Simulation (ModSim) workloads, ``hope of correctness'' is the wrong confidence level.

Existing systems are not built to support provenance by default. Through hacking things like extended POSIX attributes in the file system, some information could be added, but it would have to be explicitly handled by every client and then it may not be portable to other systems that may or not properly support these extended attributes. File format approaches all require manual application changes to write the additional information. The hardest problem to address today is how to uniquely identify an object (e.g., application, input deck, or data set) in a portable way.

Container technology characteristically offers the following features:
\begin{enumerate}
\item portability including dependent library support with a web of dependent container instances
\item isolation from other applications
\item encapsulation of a set of files into a single object (most container systems)
\item a unique hash code
\end{enumerate}

Of these attributes, the scale out community has focused on the first two. These features are ideal for offering an immutable application that can be deployed quickly without worrying about which machine it is deployed on and what the dependent libraries are. It also offers invisible scalability by trivializing adding replicas that can all offer the same service without any end-user knowledge or specific action. These features work well to address service-oriented and serverless computing. For ModSim workloads, these same characteristics are useful, but not sufficient. This aspect of using containers for ModSim is being explored in a commercial environment by companies like Rescale that offer SaaS ModSim.

Considering the generated data and the needed confidence of exactly how it was generated, the latter two attributes are the most interesting and must be explored. Encapsulating the application and input deck each into their own container is a necessary first step. By wrapping any generated data into a new container as well ensures that the data set is treated as a whole as well. With some container systems offering the ability to add annotations, it becomes possible to annotate the data, in a fully portable way, to guarantee precise knowledge of the context used to created the data. Using a system that offers these annotations enables this functionality with generally no application code changes.

This project is providing a low-level, first step for a viable fix for this problem. By encapsulating the data (and other generated artifacts) into their own system-generated container wrappers, what we call a \emph{Data Pallet}, we can use the unique ID for each container mounted (e.g., the application and input deck) as annotations directly linking the generated data to what created it. Key to this working is that it is not a dependency that would require the dependent containers be loaded in order to use this container. Then, by inspecting the Data Pallet, it is simple to find the matching container for the application and input deck using the annotation stored container IDs. There is no question of whether or not these are the matching components or not. Further, since the annotations and unique container ID are in the wrapper (i.e., not in the data itself), it is not possible to move the application, input deck, or Data Pallet without moving the IDs and annotations without explicit, difficult work. This makes identifying both the ancestry and dependency webs straightforward.

The potential impact of this work is enormous. For high consequence workloads, having an established provenance record of what source and configuration created a data set offers a guarantee we cannot ensure today. The problematic approaches being pursued both historically and today to address this issue are divorced from the underlying system that users and system processes will manipulate outside the management system causing links to get stale and broken. Only with an approach like this can we have full confidence that our ModSim analysis can be trusted and be traceable back to the original source code and people that created it. For the broader science community, reproducibility is important. To demonstrate reproducibility, a couple metrics must be met. First, data must be sufficiently traceable back to where they came from and how they were created. Second, the higher standard of replicable requires that everything is sufficiently documented to enable someone else to recreate the experiments and generate comparable result. This is what is required for something to be a science. This system enables automatic tracking and use of support information without requiring a separate tool--the information is all inherent in the artifacts themselves--making it science by default.

There are a tremendous number of aspects of this idea to explore before it can be fully exploited in systems. This first step proves the idea is worthwhile and that further exploration is warranted.

The rest of this document is divided into the following sections. First is a short look at related work in Section~\ref{sec:related}. Next, is a description of the design in Section~\ref{sec:design}. Some information about the measured overheads in this approach are included in Section~\ref{sec:measurements}. Finally, Conclusions and Future Work are described in Section~\ref{sec:conclusions}

\section{Related Work}\label{sec:related}
Existing work that attempts to do similar things does not exist. This idea and approach are wholly new. However, there are some systems and work that are doing some more distantly related things.

Workflow systems, such as the Sandia Analysis Workbench~\cite{foss:2018:saw}, Pegasus~\cite{deelman:2015:pegasus}/DAGMan~\cite{Malewicz:2006:dagman}, Kepler~\cite{bertram:2006:kepler}, Swift~\cite{wilde:2011:swift}, and many others all focus on providing a way to orchestrate a series of tasks to accomplish a goal. However, they offer little to no support for managing the artifacts, configurations for each component, and component versioning. The lack of system level support for these kinds of activities makes offering tools that provide them difficult to impossible.

The Data Pallets idea, at a surface level, is similar to various storage containers ideas. However, these approaches all focus on exposing a shared storage layer into the container context. None of them offer any sort of wrapper approach that could hold attributes linking data back to the antecedents.

Other approaches, if used correctly, have a potential to achieve the same goals, but with application modification. For example, ADIOS~\cite{lofstead:2008:adios}, NetCDF~\cite{rew:1990:netcdf}, and HDF5~\cite{koziol:1998:hdf5} all offer arbitrary attributes inside the file. If a user were to explicitly add the container IDs, they could achieve a similar goal. However, that would require that the container IDs be apparent in the application's running context and that the application code be modified to include writing such annotations into the file. The Data Pallets approach eliminates all of those requirements by adding the wrapper automatically just by virtue of running the application in a container and the annotations are generated the same way. No changes to the application or how it runs will be required.

Singularity~\cite{kurtzer:2017:singularity} offers a writeable container, but it has some strict limitations. First, the container has to be created prior to running Singularity and mounted as part of the startup process with an option to indicate that it should be writeable. Rather than limit to a single container, manually created by the user prior to starting the application, we are proposing to auto-generate and mount new writeable containers at runtime. Further, we will annotate these generated containers with the IDs from the context in which it was created. This is the key traceability component.

\section{Design}\label{sec:design}
The design for this idea is very simple. Intercept any attempt to create a file on storage and create and mount a new container, with the appropriate annotations applied, to capture that output. While this simple idea is currently not fully attainable for a variety of reasons, there are ways to work around these limitations to demonstrate the idea potential. Further, for this to be a workable approach, it needs to work within a production workflow system, such as the Sandia Analysis Workbench.

\begin{figure}[!h]
\centering
 \includegraphics[width=0.5\textwidth]{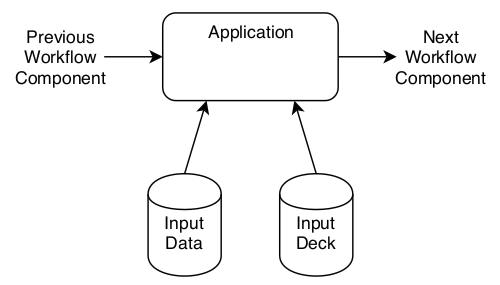}
 \caption{Typical ModSim Usage}
 \label{fig:workflow-1}
\end{figure}

To visualize how the Data Pallets work, consider the progression of these figures. Figure~\ref{fig:workflow-1} is an illustration of what a typical workflow component looks like. It is a node in a graph with an application using input data from somewhere, typically storage, and an input deck that describes how to operate.

\begin{figure}[!h]
\centering
 \includegraphics[width=0.5\textwidth]{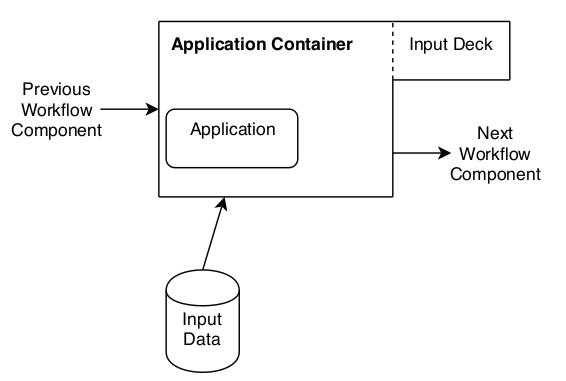}
 \caption{Containerized ModSim Usage}
 \label{fig:workflow-2}
\end{figure}

Figure~\ref{fig:workflow-2} shows how this evolves if containers are introduced. Note that the application is wrapped in a container and the input deck is somehow mounted into that container. It may be via mounting a host directory into the container, or as illustrated, as a separate container that is mounted into the same namespace. Most importantly, the input data comes from storage, or other sources, still.

\begin{figure}[!h]
\centering
 \includegraphics[width=0.5\textwidth]{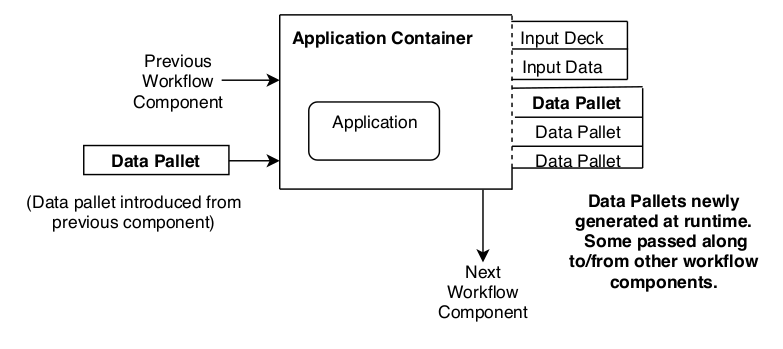}
 \caption{Containerized \& Palletized ModSim Usage}
 \label{fig:workflow-3}
\end{figure}

The final version shown in Figure~\ref{fig:workflow-3} expands the containerized version to shift input data from being read from some external source to being brought in via either a container or a native mount. We also illustrate the dynamically created Data Pallet containers that are also mounted into the namespace. However, note that the first Data Pallet is potentially a Data Pallet created by a previous component. In this case, the Pallet would likely be read only. Making it writeable may lead to confusion about the data provenance. At a minimum, the annotations would need to be expanded to include the new container context.

With this new architecture, containers proliferate along with the provenance annotations. This is not a problem-free approach, but it solves the data provenance issue. The implications of this approach are many. Some of them are explored in this paper while others are being better formulated and explored still before being shared.

\subsection{Design and Implementation Challenges}
The way that containers work makes enabling this approach more difficult than it may appear. Containers are a way to work with the Linux Namespaces~\cite{biederman:2006:linux-namespaces} easing setup and use. In particular, what happens is that the container system creates a namespace and maps in a series of resources. All other resources on the machine are not accessible. This enables having multiple instances running in isolation from each other with no changes necessary in the application.

The implication of this namespace implementation is that there are no intermediaries between a containerized application and the resources, such as a translation layer. This enables essentially native performance. The difficulty is that it does not enable experimenting with new system-level functionality, such as that proposed with Data Pallets. Instead, it is still necessary to use standard OS implementation techniques to introduce new features. To achieve the Data Pallets idea, a classic file systems research approach was employed.

The initial implementation idea was to intercept a call to the system command \texttt{mkdir}, the command to create a directory in storage, and use that as the entry point for creating a new container on the fly and mounting it. Since the container system did not offer an intercept point and instead exposed the kernel directly, FUSE~\cite{szeredi:2005:fuse} was a natural way to intercept the system call invoking the functionality.

Long term, a system level approach for intercepting POSIX \texttt{open} and other file or storage object creation calls would offer a truly native way to accomplish the same goal.

\subsection{Design and Implementation Details}
To accelerate the idea exploration, we chose to leverage the Singularity system. It already offered a way to mount an existing container as ``writeable'', but did not offer a way to generate that on-demand nor to annotate it with any sort of context information. 

To make this work, we used Singularity 3.0 for two main reasons. First, Singularity's security model offers the best solution for working in the sensitive DOE environment. Docker's passing root access from inside a container to the underlying system is not viable for highly secured systems. Singularity's approach isolates root access in a container to just the container, preserving the security envelope. Second, Singularity introduced enhanced a new container image format that supports multiple streams (partitions) making adding annotations easier. In our case, we add a JSON file with the annotations.

\subsection{Integration with Sandia Analysis Workbench (SAW)}
Integrating with SAW demonstrates that the Data Pallets idea is compatible with production workflow systems typically used to run ModSim workloads. Making Data Pallets work within SAW required no special changes for the most basic level. Instead, a new object type to handle a container was the only requirement. This capability is a standard end-user enabled feature part of SAW. A richer integration, such as offering an interface to the container hub making selecting an application or inspecting annotations easier, are beyond the scope of this work.

\section{Measurements}\label{sec:measurements}
To evaluate the implementation, tests are performed on a single desktop workstation. While this is not an ``at scale'' implementation, it is sufficient to show how this can work for file per process workloads. Alternative workloads that use an N-1 pattern will require further exploration.

The testing environment is a Dell Precision 3420 SFF tower. It has an Intel Core i7-6700 CPU @ 3.40 GHz, 32 GB RAM, and a 256 GB ATA SAMSUNG SSD SM87 for storage. It is running Ubuntu 16.04 LTS - 64-bit. The Singularity version is the initial release of 3.0. FUSE 3.3.0 is used.

\subsection{Time Overheads}
The time overheads for this approach are small, particularly in the context of the application runtimes. To evaluate the time, we use a simple application, gnuplot, and make 1000 tests using the mean of the results for each test.

To determine the overheads, we evaluate the following characteristics. First, we measure the time it takes for the application to run generating output completely outside of the container environment. Second, we measure the time it takes to run the application from within a container, but writing directly to storage. Third, we measure the time with the application in a container writing into a dynamically created container. Using these three metrics, we are able to extract these detailed overheads.

In Table~\ref{tab:workflow-node-overhead}, we show the metrics related to running an application, gnuplot, using the structure described in Figure~\ref{fig:workflow-3}. By far, the time is spent in the container load overhead. Considering the time to deploy applications in a cluster, around 0.5 seconds is negligible. For cloud application deployments, this overhead is typical and considered sufficiently fast to be useful for a production environment.

Table~\ref{tab:application-containers-overhead} shows the time to create the various components. As is shown, the overheads are tiny and acceptably small considering the variability in any parallel storage system. This would be completely lost in the noise.

\begin{table*}[ht]
\centering
\caption[Workflow Node Overheads]{Workflow Node Overheads}
\bigskip

\begin{tabular}{|l|l|}
\hline
Average time to mount the custom FUSE filesystem & 0.037 sec\\
\hline
Average time to spin up / tear down Application Container & 0.498 sec\\
\hline
Average time to run application and create Data Pallets & 0.037 sec\\
\hline
Average total time to run the Application Container & 0.535 sec\\
\hline
Average time to unmount the custom FUSE filesystem & 0.029 sec\\
\hline
Total time taken by the Gnuplot Workflow Node & 0.601 sec\\
\hline
\end{tabular}
\label{tab:workflow-node-overhead}
\end{table*}

\begin{table*}[ht]
\centering
\caption[Application Containers Overheads]{Application Containers Overheads}
\bigskip

\begin{tabular}{|l|l|}
\hline
Average time to run Gnuplot on input data & 0.0284 sec\\
\hline
Average time for total Data Pallet creation / storage & 0.00133 sec\\
\hline
Average time to pull PNG output file from Data Pallet & 0.00725 sec\\
\hline
\end{tabular}
\label{tab:application-containers-overhead}
\end{table*}

\subsection{Space Overheads}
Since we are incorporating additional data into each output and requiring similar overheads on applications and input decks, it is important to understand the space overheads imposed by this approach. To determine these overheads, we test the following configurations:
\begin{itemize}
    \item Empty Container - This is the baseline overhead for all applications and input decks. Since we will store all of these artifacts long term, understanding this base overhead will reveal how much additional storage will be required for archiving.
    
    \item Basic Writeable Container - This is the size overhead for the generated data sets.
    
    \item Attributes Stream - Since each container image can contain multiple data streams, one will be the normal data contents and a second will be the JSON file containing the list of associated container IDs.
\end{itemize}

	\begin{table}[ht]
	    \centering
	    \caption[Space Overheads]{Space Overheads}
	    \bigskip

	    \begin{tabular}{|l|l|}
	    \hline
	    Empty Container & 32.2 KB\\
	    \hline
	    Writeable Container & 704.5 KB\\
	    \hline
	    Attributes Stream & 1.1 MB\\
	    \hline
	    \end{tabular}
	    \label{tab:space-overhead}
	\end{table}

The space overheads are shown in Table~\ref{tab:space-overhead}.

\subsection{Discussion}
From this data, it is apparent that the primary source of overhead in running the workflow node is the application container, itself. The time by the Data Pallets creation and management is minimal as compared to the application runtime. The total time taken by the internals of the container are also minimal as compared to the spin up and tear down time of the container itself. Fortunately, the overhead involved in container creation is a prevalent issue in current research, and strides are being made to minimize it. Out of the overhead introduced by the new approach, the majority of increase comes from the mounting of the FUSE filesystem. Actual use of it to create the SIF images is negligible.

The empty container size is small enough to be used as a wrapper around any application. For an input deck, this size is likely larger than the input deck itself (excepting a mesh description), but still small enough to justify using it to wrap around the input deck.

The writeable container size is much larger because the only writeable format is ext3 rather than the more compact squashfs, what is used for read-only containers. The reason ext3 is required is that the container system requires the underlying operating system to handle the writing operations into storage backed container file. The squashfs file system format inherently does not support writing into it except at creation.

The attributes stream overhead is surprisingly large. This would be the minimum size for each additional attribute stream added to a container image. For output data sets on the order of 10\% of node memory, this is still a small enough overhead to be acceptable. Our nodes are moving to 128 GB as the base memory making this less than 1\% of the total volume written during the output.

\section{Integration With Sandia Analysis Workbench}\label{sec:measure-saw}
The long-term goal for this work is to be able to deploy fully container wrapped applications and input decks into the SAW system and eliminate the need for SAW to attempt to track generated artifacts and instead rely on the inherent annotations as the source of truth. The previously mentioned limitations with the current system that attempts to track artifacts without owning them is fragile with many exposed areas where the system can be broken either accidentally, on purpose, or by automated system processes.

To test this integration, a simple input deck was wrapped in a container, another container contained an application, and it generated a separate output. We chose to use the gnuplot application as there is a simple example using this provided as part of SAW.

By wrapping everything in a container, we changed the workflow to run the component by using the container runtime command rather than directly running the application. This simple change revealed the simplicity with which this change could be deployed with minimal to no impact on existing users. Further integration is left to future work.

\section{Conclusions and Future Work}\label{sec:conclusions}
This work demonstrates that it is possible to shift downward the provenance information necessary to trace from a given data set back to the application and input deck that created it. Unlike other approaches, Data Pallets offers a guarantee of this traceability without having to use a special tool for all data access. Instead, by incorporating the data annotations linking the data back to the application and input deck, the provenance is manifest rather than linked. The potential benefit for both reproducibility and confidence in ModSim results is enormous.

For the future work, many steps will be investigated.

First, considering current limitations in how Singularity mounts containers, we can create a container at runtime, but it only works properly when we add the files to the newly created container after the files are created. Sylabs and Sandia are currently working together to change how things work to allow files to be written to the dynamically created container.

Second, issues surrounding when and how to unmount a writeable container need to be investigated. Currently, we are assuming that the system can flush cold data to storage without having to rely on the container context exiting. If this is not the case, then memory pressures are going to force changes such that we can discover how to unmount the created container to free needed memory.

Third, the SAW integration demonstrated the possibility of running containers. The next step for this to be usable and viable is to offer introspection into things like the JSON attributes stream, seeing a list of dependent containers, and getting a list of available containers to run from a local container ``hub''. These hubs offer a way to dynamically load a container as needed based on the container ID rather than having to manually pre-deploy all needed containers when running the application. Scalability issues with this are likely at the high end, but well understood and demonstrated techniques to manage this, such as distribution trees, can address the performance issues. Further, the current Singularity container hub uses Docker's system, something not viable within our environment.

\section*{Acknowledgements}
Sandia National Laboratories is a multimission laboratory managed and operated by National Technology and Engineering Solutions of Sandia, LLC, a wholly owned subsidiary of Honeywell International, Inc., for the U.S. Department of Energy's National Nuclear Security Administration under contract DE-NA0003525. 
This work is funded through the LDRD program and ASC CSSE.

\bibliographystyle{plain}
\bibliography{sample-bibliography}

\end{document}